\documentclass[aps,prb,10pt,twocolumn]{revtex4-1}
\usepackage{amsmath}
\usepackage{graphicx}
\usepackage{amssymb}

\usepackage{gt17}
\usepackage{color} 
\usepackage{graphicx}
\usepackage{bm}
\allowdisplaybreaks 
\begin{document}
\title{
Pair emission from a relativistic domain wall in antiferromagnets
}
\author{Gen Tatara$^{1,2}$} 
\author{Collins  Ashu  Akosa$^{1,2}$}
\author{Rubén M. Otxoa de Zuazola$^{3,4}$}
\affiliation{$^1$RIKEN Center for Emergent Matter Science (CEMS)
and RIKEN Cluster for Pioneering Research (CPR), 
2-1 Hirosawa, Wako, Saitama, 351-0198 Japan}
\affiliation{$^2$ Department of Theoretical and Applied Physics, African University of Science and Technology (AUST), Km 10 Airport Road, 
Galadimawa, Abuja F.C.T, Nigeria}
\affiliation{$^3$Hitachi Cambridge Laboratory, J. J. Thomson Avenue, CB3 OHE, Cambridge, United Kingdom}
\affiliation{$^4$ Donostia International Physics Center,  20018 San Sebasti\'an, Spain}
\date{\today}

\begin{abstract}
Magnon emission and excitation by a relativistic domain wall at a constant velocity in antiferromagnet is theoretically studied.
A pair emission due to a quadratic magnon coupling is shown to be dominant. The emission corresponds in the comoving frame to a vacuum polarization induced by a zero-energy instability of the Lorentz-boosted anomalous response function. 
The emission rate is sensitive to the magnon dispersion and wall profile, and is significantly enhanced for a thin wall with velocity close to the effective light velocity. 
The Ohmic damping constant due to magnon excitation at low velocity is calculated.
\end{abstract}  

\maketitle
\newcommand{\Vw}{v_{\rm w}}

Emission from a relativistic moving object is a general intriguing issue that has analogy to blackbody radiation, blackhole physics \cite{Petev13} and can be applied for wave amplification \cite{Ostrovskii72}.
Solid-state systems are particularly interesting from the viewpoints of quantum effects and experimental feasibility due to low 'light velocity'. 
Antiferromagnets at low energy have been known to be typical relativistic system \cite{Haldane83}, and dynamic properties of domain wall has been explained in terms of Lorentz contraction \cite{Shiino16}. 

In this paper, we study the emission from moving domain wall, a relativistic soliton, in an antiferromagnet.
We discuss the low energy regime using a continuum model, valid when the wall thickness $\lambda$ is larger than the lattice constant $a$.  
The system is described by a relativistic Lagrangian, and thus there are domain wall solutions moving with a constant velocity smaller than the effective light velocity $c$.
The wall width $\lambda$ is affected by Lorentz contraction; $\lambda=\gamma \lambda_0$, where $\lambda_0$ is the thickness at rest,  
$\gamma(\Vw)\equiv \sqrt{1-(\Vw/c)^2}$ is a contraction factor, $\Vw$ is the velocity of the wall.

Emission from a moving object is generally dominated by a linear process, where the object couples to its fluctuation linearly. 
In the case of soliton solutions, such linear coupling, absent at rest, arise from acceleration and deformation as argued for ferromagnetic domain wall \cite{Bouzidi90,LeMaho09,KimSW18,TataraSW20}. 
The antiferroamgnetic case turns out to be qualitatively distinct from the ferromagnetic case because of the Lorentz invariance. 
The linear coupling, inducing super Ohmic dissipation, is negligible at low energy, and the dominant emission arises from the second-order coupling to the moving wall. The momentum is transferred from the wall to magnons, while the energy comes from Doppler shift. 
In the rest frame of the wall, the wall potential generates a localized magnon excitation.
The excitation is described by the normal (particle-hole) component of magnon response function, which we call $\Pi_q$ ($q$ is the wave vector transferred). 
In the moving frame, this excitation corresponds to a scattering of magnon, resulting in an Ohmic friction force at low velocity. 
The scattering property of the normal response function $\Pi_q$ is essentially the same as in the ferromagnetic case studied in Ref. \cite{TataraSW20}; 
Although the magnon dispersion in ferromagnet, quadratic in the wave vector $k$, is different from the antiferromagnetic linear behavior (in the absence of gap), it does not lead to qualitative difference in magnon scattering.

A significant feature antiferromagnets have is the existence of an anomalous particle-particle (or hole-hole) propagation, $\Gamma_q$, like in superconductivity contributing to the response function \cite{TataraAF19}. 
This is due to the quadratic time-derivative term of the relativistic Lagrangian, which allows  positive and negative energy (or frequency) equally.
The anomalous response function thus can be regarded as a scattering of particles with a positive and negative energies. 
The negative frequency mode exists generally in any relativistic excitations.
In optics, for example, a scattering of negative frequency mode was argued to cause an amplification of photon current \cite{Rubino12}.
In the context of magnons, the scattering of negative frequency mode corresponds to an emission/absorption of two magnons.
The anomalous response function $\Gamma_q$ describing such process is shown to be sensitive to the magnon dispersion as well as the wall velocity.   
Its low energy weight is much smaller compared to the normal response function $\Pi_q$ for the ideally relativistic dispersion of $k$-linear dependence, while it is significantly enhanced if it deviates from linear to have a flatter dispersion. 
The anomalous response function in this case has a sharp and large peak at finite wave vector for the wall velocity close to the effective light velocity $c$, resulting in a strong forward emission of two magnons.
Our results indicates that relativistic domain wall is useful as  a magnon emitter, and the efficiency is tunable by designing magnon dispersion.

The pair emission here is analogous to the vacuum polarization (Schwinger pair production) in electromagnetism \cite{Schwinger51}, with the role of electric field played by the moving wall.
In fact, in the laboratory frame, the magnon creation gap of $2\Delta$ is overcome by the energy shift by the Doppler's effect, while in the moving frame with the wall, a spontaneous vacuum polarization is induced by a zero-energy instability of the Lorentz-boosted anomalous magnon response function.


Magnetic properties of antifferomagnets are described by the staggered (N\'eel) order parameter $\nv$ of the unit length. 
Its low energy Lagrangian is relativistic, namely, invariant under the Lorentz transformation as for the kinetic parts \cite{Haldane83}. 
We consider the case with an easy axis anisotropy energy along the $z$ axis, described by the continuum Lagrangian of 
\begin{align}
L & = \frac{J}{2a}\int dx \lt[ \frac{1}{c^2}\dot{\nv}^2 -(\nabla \nv)^2 +\frac{1}{\lambda_0^2}(n_z)^2\rt],
\end{align}
where $J$ is the exchange energy, $J/\lambda_0^2(=K)$ is the easy axis anisotropy energy.
Our results are valid in the presence of hard-axis anisotropy simply by including the effect in the gap of magnons. 
The effective light velocity is $c=\sqrt{gJ}$, $g$ being a coupling constant \cite{TataraAF19}.
The lattice constant is included to simplify the dimensions of material constants \cite{AFSW20Yang}.  
We consider the one-dimensional case, although the the effects we discuss are general and apply to higher-dimensional walls. 
The Lagrangian is relativistic, i.e.,  a Lorentz transformation to a moving frame with a constant velocity $v$, 
 $t'= ({t-\frac{v}{c^2}x})/\gamma(v)$ and  $x'= ({x-{v}t})/\gamma(v)$ 
does not modify the form.
The system has a soliton (domain wall) solution, $n_z(x)=\tanh \frac{x}{\lambda_0}$.
The Lorentz invariance indicates moving walls $n_z((x-vt)/\gamma)$ are classical solutions for a constant $v<c$, with a contracted thickness $\lambda=\lambda_0\gamma(v)$. 

These constant velocity solutions are stable, meaning that they have no linear coupling to magnons and there is no linear emission.
Linear emission may occur during acceleration or by deformation.  
The emission is studied by introducing collective coordinates \cite{TKS_PR08}. 
In the case of a domain wall, of most interest is the wall position $X$ \cite{AFSW20othercollective}. 
The coupling between the coordinate and fluctuation is governed by the kinetic part of the Lagrangian.
In antiferromagnets, it is second order in time derivative, and thus linear fluctuation, $\varphi$, couples to the acceleration $\ddot{X}$ as $\varphi\ddot{X}$ (See Ref. \cite{TataraSW20}).
The emitted magnon amplitude $\average{\varphi}$ is thus proportional to $\ddot{X}$, and the recoil force on the wall  is  
$\frac{\partial^2}{\partial t^2}\average{\varphi}\propto \frac{\partial^4}{\partial t^4}{X}$. 
Hence the linear coupling does not induce Ohmic friction and is negligibly small at low energy. 
The result is the same for other collective variables like thickness oscillation.  
The motion of an antiferromagnetic domain wall is therefore protected from the damping due to a linear coupling, in contrast to the ferromagnetic case, where Ohmic dissipation arises from thickness oscillation \cite{TataraSW20}.

\begin{figure}
 \includegraphics[width=0.6\hsize]{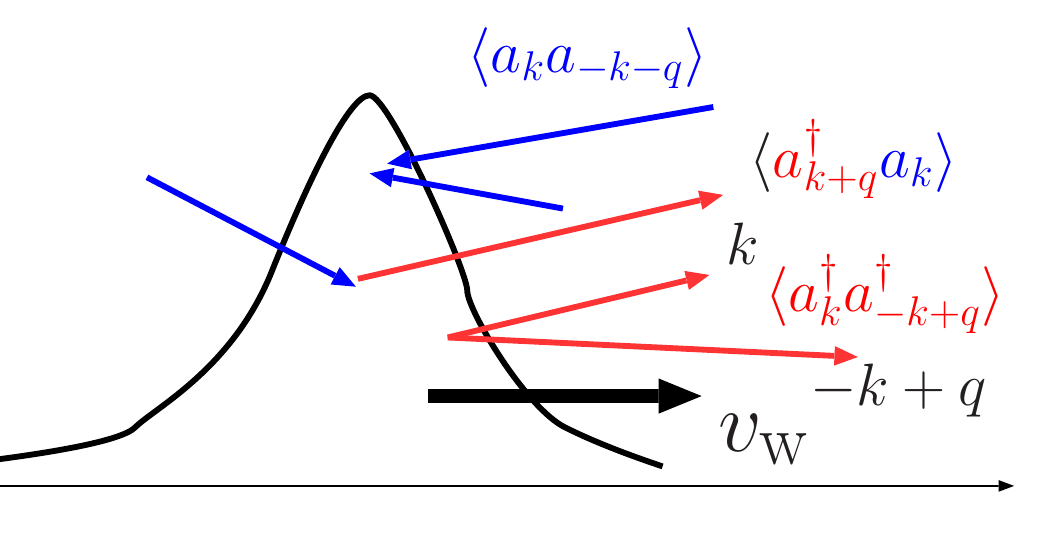}
 \caption{
 Schematic figure showing processes due to moving domain wall, scattering $\average{ a^\dagger a}$, pair emission $\average{a^\dagger a^\dagger}$ and pair annihilation, $\average{aa}$. Momentum $q$ is transferred from the wall with a velocity $\Vw$ to the magnons. 
 \label{FIGdwemission}}
\end{figure}

Instead, emission due to the second-order coupling dominates in antiferromagnets.
At low energy, contribution containing less time derivative of the wall collective coordinates dominates.
The issue then reduces to a simple and general problem of the emission from a moving potential of a constant velocity \cite{AFSW20Kim}.
Our domain wall solution of $\tanh$-profile induces an attractive potential of $\cosh^{-2}$ form.
Taking account of the two magnon modes along the $x$ and $y$-directions, $\varphi_x$ and $\varphi_y$, respectively 
($\nv\simeq (\varphi_x,\varphi_y, 1)$), the potential reads \cite{TKS_PR08} 
\begin{align}
V&=  -K\int \frac{dx}{a} \frac{1}{\cosh^2 \frac{x-X(t)}{\lambda}}(\varphi_x^2+\varphi_y^2),
\label{Vvarphi}
\end{align}
where $X(t)$ is the wall position and $\lambda=\lambda_0\gamma$ is the thickness of a moving wall \cite{AFSW20Tveten}. 
We consider the case of a constant velocity, $X(t)=\Vw t$.
A moving potential transfers momentum $q$ to fluctuations and  an angular frequency $\Omega$ as a result of  the Doppler shift.
Although the form of the potential, Eq. (\ref{Vvarphi}), is common for ferro and antiferromagnetic cases, its effect is different, due to different nature of magnon excitations.
In ferromagnets, $\varphi_x$ and $\varphi_y$ are represented as linear combination of magnon field $b$ and $b^\dagger$ (Holstein-Primakov boson).   
The potential in this case is proportional to magnon density as $\varphi_x^2+\varphi_y^2= 4 b^\dagger b$, inducing scattering of magnons without changing total magnon number.
(The feature is unchanged in the presence of a hard-axis anisotropy.)
This is due to the kinetic term linear in the time-derivative for ferromagnetic magnon \cite{TKS_PR08}, $ib^\dagger \dot{b}$, which allows a positive energy for the ferromagnetic magnon boson.
In contrast, a magnon boson in antiferromagnets is described by a relativistic Lagrangian 
with a kinetic term second-order of time derivative,  $\frac{1}{c^2} (\dot{\varphi}_i)^2$ ($i=x,y$), which allows 'negative frequency' modes , and processes changing the  total magnon number are allowed. 
In fact, canonical magnon boson $a_i$ is defined for each mode $i=x,y$ as 
$\varphi_i(k,t) =\sqrt{\frac{g}{\omega_k^{(i)}}} ( a^{(i)}_k(t)+a^{(i)\dagger}_{-k}(t))$, where 
$\omega_k^{(i)}\equiv \sqrt{c^2k^2+\Delta_{i}^2}$ is the energy with a gap $\Delta_i$ of mode $i$ \cite{TataraAF19}.  
The potential, Eq. (\ref{Vvarphi}), then reads  
\begin{align}
V =&   -Kg \frac{\lambda}{a} \sum_{i=x,y} \sum_{k,q} \frac{W_q }{\sqrt{\omega_k^{(i)}\omega_{k+q}^{(i)}} } e^{-iqX(t)} 
 \nnr& \times 
 \lt(a_k^{(i)} a_{-k-q}^{(i)}+a_{-k}^{(i)\dagger} a_{k+q}^{(i)\dagger}
 +2a_{k+q}^{(i)\dagger}a_k^{(i)}\rt),
 \label{Va}
\end{align}
where 
$W_{q}=\pi\frac{q\lambda}{\sinh \frac{\pi}{2}q\lambda}$ is the Fourier transform of the potential profile and  $\omega_{-k}=\omega_k$ is assumed.
The emission and absorption of two magnons, represented by terms $aa$ and $a^\dagger a^\dagger$,  are thus possible in antiferromagnet (Fig. \ref{FIGdwemission}).  

\begin{figure}
 \includegraphics[width=0.6\hsize]{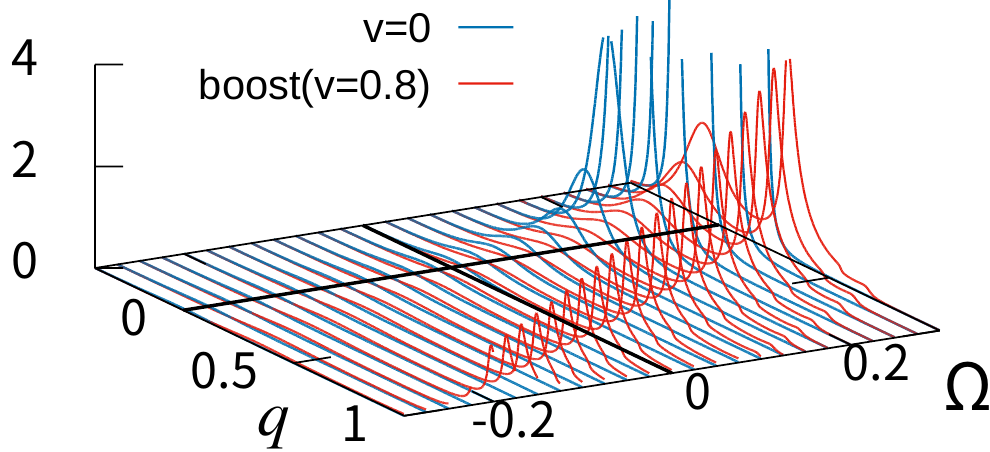}
 \includegraphics[width=0.38\hsize]{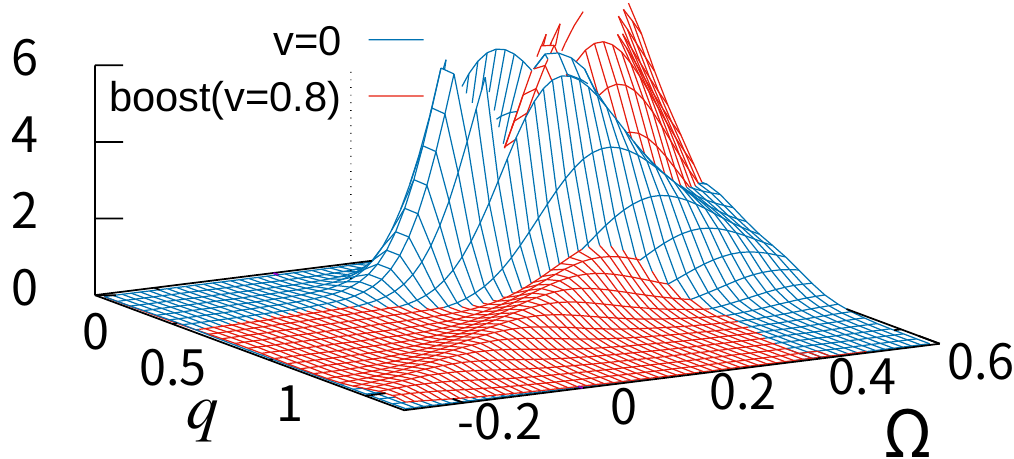}
 \caption{Effect of Lorentz boost on the anomalous response function $\Im[\Gamma_{q,\Omega}]$ at $\Vw/c=0.8$ for a hyperbolic dispersion ($\mu=5$), $\tilde{\lambda_0}=2$, $\tilde{\Delta}=0.1$, $\tilde{\eta}=0.01$ and $\tilde{T}=0.2$ in dimensionless unit (See Ref. \cite{AFSWSM20}). Blue is the amplitude at rest frame, which is localized at $\Omega \gtrsim 2{\Delta}$ with negligibly small weight at $\Omega=0$. In the boosted frame, shown in red, the amplitude extends to zero energy regime at finite $q$, inducing spontaneous vacuum polarization, which corresponds to a pair emission in the laboratory frame. 
 \label{FIGIqboost}}
\end{figure}

Let us evaluate the amplitudes of scattering and emission/absorption as a linear response to the dynamic potential.
We suppress the index $i$ for magnon branch. 
The scattering amplitude, $\average{ a_{k+q}^{\dagger} a_k }(t)=iG^<_{k,k+q}(t,t)$, is a lesser Green's function for magnon.
The amplitude after summation over $k$ is represented in terms of  the normal (particle-hole) response function (including the form factor $W_q$)\cite{TataraSW20,TataraAF19},  
\begin{align}
\Pi_{q,\Omega}& \equiv
 - \sum_k \frac{W_{q}}{\sqrt{\omega_k\omega_{k+q}} } 
\frac{n_{k+q}-n_k}{\omega_{k+q}-\omega_{k}-\Omega+2i\eta} ,
\end{align}
as 
$\sum_k \average{ a_{k+q}^{\dagger} a_k }=\frac{Kg}{a}  \lambda e^{iq\Vw t} \Pi_{q}$, where $\Pi_q\equiv \Pi_{q,q\Vw}$. 
Here $n_k\equiv [e^{\beta\omega_k}-1]^{-1}$ is the Bose distribution function, 
$\beta\equiv (\kb T)^{-1}$ being the inverse temperature ($\kb$ is the Boltzmann constant),  $\eta$ is the damping coefficient of the magnon Green's function.
The angular frequency of $\Omega=q\Vw$ in $\Pi_{q,q\Vw}$  is the one transferred to magnons as a result of the Doppler shift.  
The  emission amplitude of two magnons is 
$\sum_k \average{ a^{\dagger}_{-k+q} a^{\dagger}_{k} } = \frac{Kg}{a} \lambda 
e^{iq\Vw t}   \Gamma_{q}$, where  $\Gamma_{q}\equiv \Gamma_{q,q\Vw}$ and 
\begin{align}
\Gamma_{q,\Omega} 
&\equiv \sum_k 
  \frac{W_{q} }{\sqrt{\omega_k\omega_{-k+q}} }
  \frac{1+n_{-k}+n_{-k+q} }{\omega_{-k+q}+\omega_{k}-\Omega +2i\eta} ,
\end{align}
is the anomalous (particle-particle) response function.
The absorption amplitude is given by this function as 
$\sum_k \average{ a_{-k+q} a_{k} } = \frac{Kg}{a} \lambda e^{iq\Vw t}   \Gamma^*_{-q}$ ($^*$ denotes the complex conjugate). 
The normal response function has symmetry of $\Pi_{-q,\Omega}=\Pi_{q,\Omega}$, which leads in the case of $\Omega=q\Vw$ to 
$\Pi_{-q,-q\Vw}=\Pi_{q,q\Vw}^*$, i.e., the real (imaginary) part of $\Pi_q$ is even (odd) in $q$. 
The normal response has low energy contribution around $q=0$ and $\Omega=0$. 
The asymmetric  and localized character near $q=0$ of $\Im[\Pi_{q}]$ \cite{AFSWSM20} indicates an asymmetric real-space magnon distribution with respect to the wall center similarly to the ferromagnetic case \cite{TataraSW20}.
The anomalous response satisfies $\Gamma_{-q,\Omega} =\Gamma_{q,\Omega}$.
It has a gap of $2\Delta$ for $\Omega$, suppressing the low energy contribution in the rest frame (Fig. \ref{FIGIqboost}).
In the moving frame, the Lorentz boost, which transforms $q$ and $\Omega$ to $q'=(q+\Vw\Omega/c^2)/\gamma$ and  $\Omega'=(\Omega+\Vw q)/\gamma$,  distorts the response function, enhancing significantly the low energy weights at finite $q$.   
This induces spontaneous vacuum polarization, which corresponds to a 2 magnon emission in the laboratory frame.

There are two key factors governing the response functions, the form factor $W_q$ and magnon dispersion.
The form factor constrains the wave vector transfer $q$ to $|q|\lesssim \lambda^{-1}=(\lambda_0\gamma)^{-1}$. 
Because of this factor,  magnon effects are significantly enhanced for thin walls at high velocity (small $\gamma$). 
As the emission is dominated by the large $q$ behavior, it is sensitive to the wall profile as we shall see below.

\begin{figure}
 \includegraphics[width=0.4\hsize]{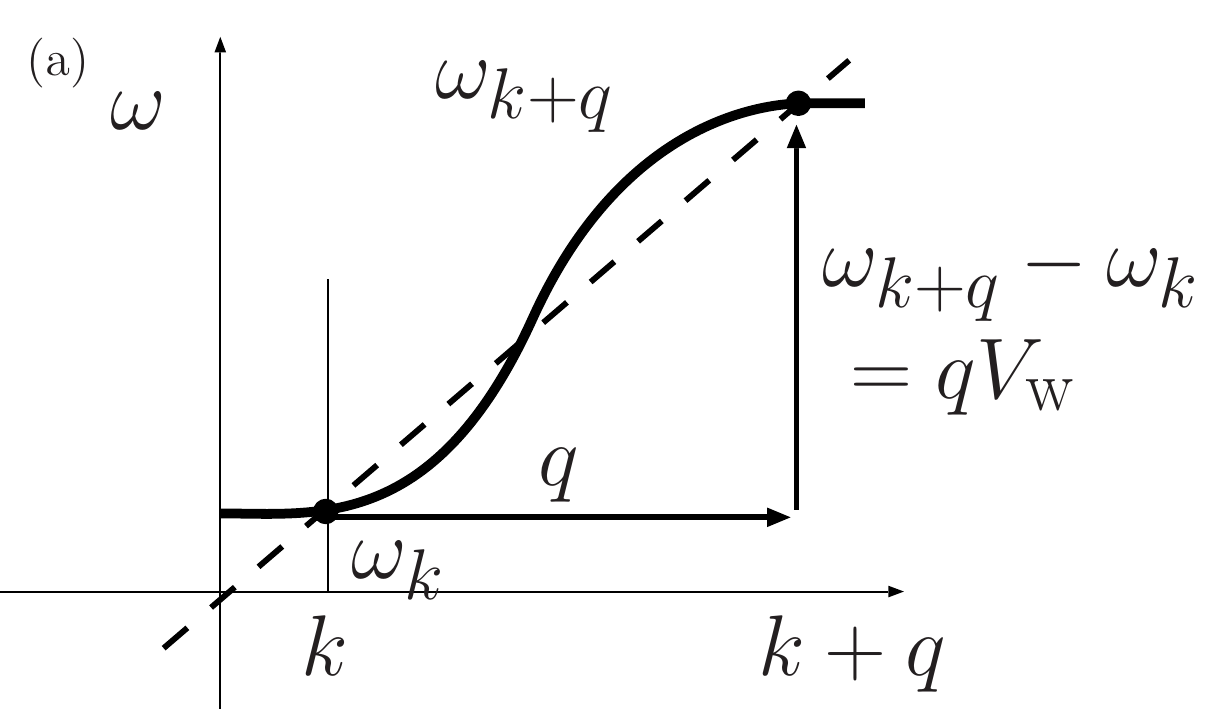}
 \includegraphics[width=0.4\hsize]{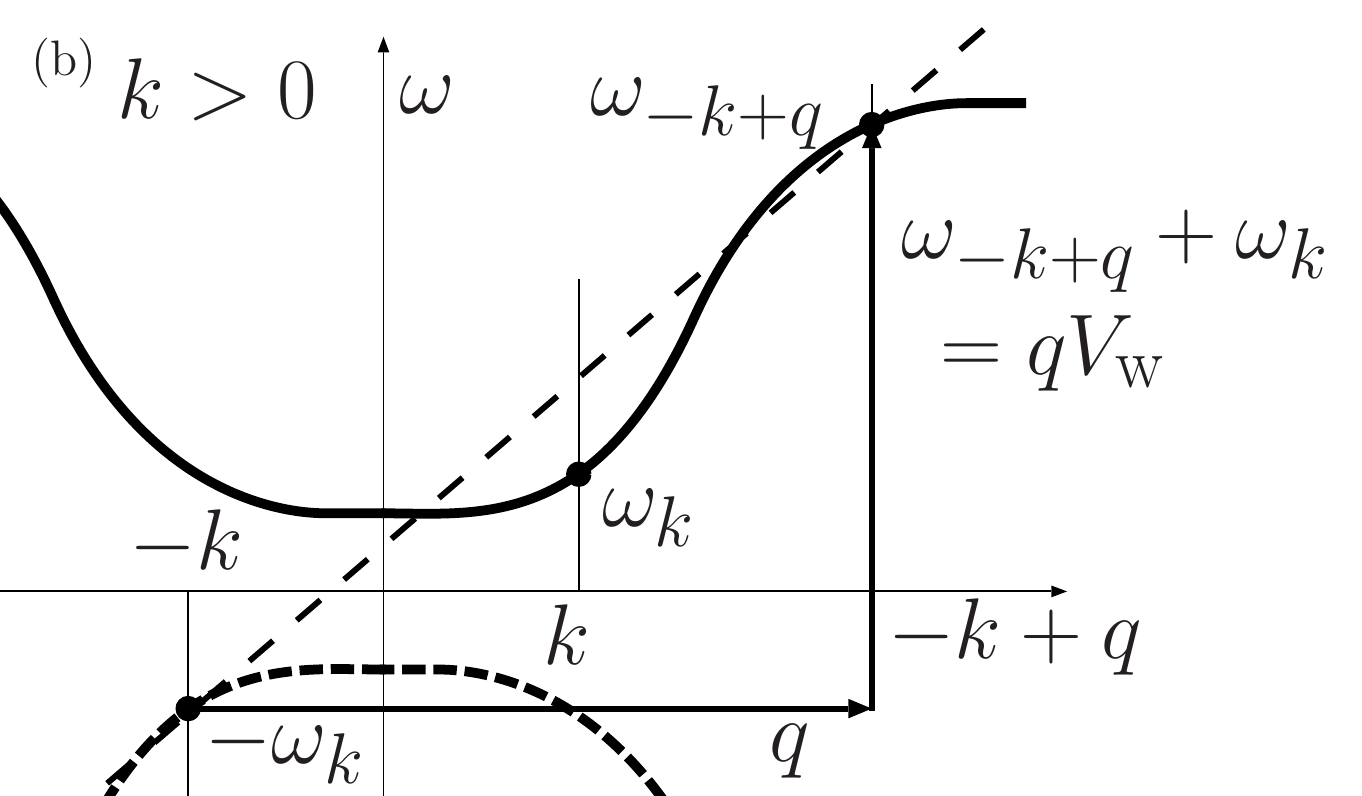}
 \caption{ Energy conservation conditions in (a) scattering and (b) emission/absorption of  two magnons.
 The emission is regarded as a scattering from a hole state with energy $-\omega_k$ to a particle state with energy $\omega_{-k+q}$. 
 The slope of  dotted straight lines is $\Vw$. 
 \label{FIGenergyconserv}}
\end{figure}

The role of the dispersion is clearly seen focusing on the imaginary part in the limit of $\eta\ra0$, where the response aries from the processes satisfying the energy and momentum conservation.
We consider the case of the dispersion with a small gap and saturation around $k_{\rm max}=\pi/a$ (See Ref. \cite{AFSWSM20}), like the one in MnF$_2$ \cite{Rezende19}. 
We choose $\Vw$ as positive.
The imaginary part of the normal response arises from the process satisfying 
 $\omega_{k+q}-\omega_{k}=q\Vw$ (Fig. \ref{FIGenergyconserv}(a)), 
which leads to an asymmetric weight around $q=0$. 
The imaginary part of the anomalous response $\Gamma_{q}$ arises when   
\begin{align}
 \omega_{-k+q}+\omega_{k} &=q\Vw .
 \label{energyconsemission}
\end{align}
This amplitude is much smaller than the normal response for the  relativistic dispersion, $\omega_k=\sqrt{(ck)^2+\Delta^2}$, due to the following reason 
(Fig.  \ref{FIGenergyconserv}(b)). 
The process satisfying Eq. (\ref{energyconsemission}) is regarded as a scattering process of a particle and a hole having positive and negative energy,
 $\omega_{-k+q}$ and $-\omega_{k}$, respectively.
The condition requires that the average slope of the line connecting the two energies $\omega_{-k+q}$ and $-\omega_{k}$ is $\Vw$. 
However, the slope is larger than $c$ for the relativistic dispersion, while 
 $\Vw$ has an upper limit of $c$, which is the maximum group velocity.
The condition cannot therefore be satisfied by the purely relativistic dispersion, and the imaginary part of anomalous response thus arises only if the dispersion has an inflection point like in   Fig.  \ref{FIGenergyconserv}(b).
(In reality, a damping $\eta$ leads to a finite imaginary part, but it remains to be  negligibly small.)
Those features are consistent with a theory of spin transport in antiferromagnet \cite{TataraAF19} showing that the anomalous correlation function is negligible.  

\begin{figure}
 \includegraphics[width=0.49\hsize]{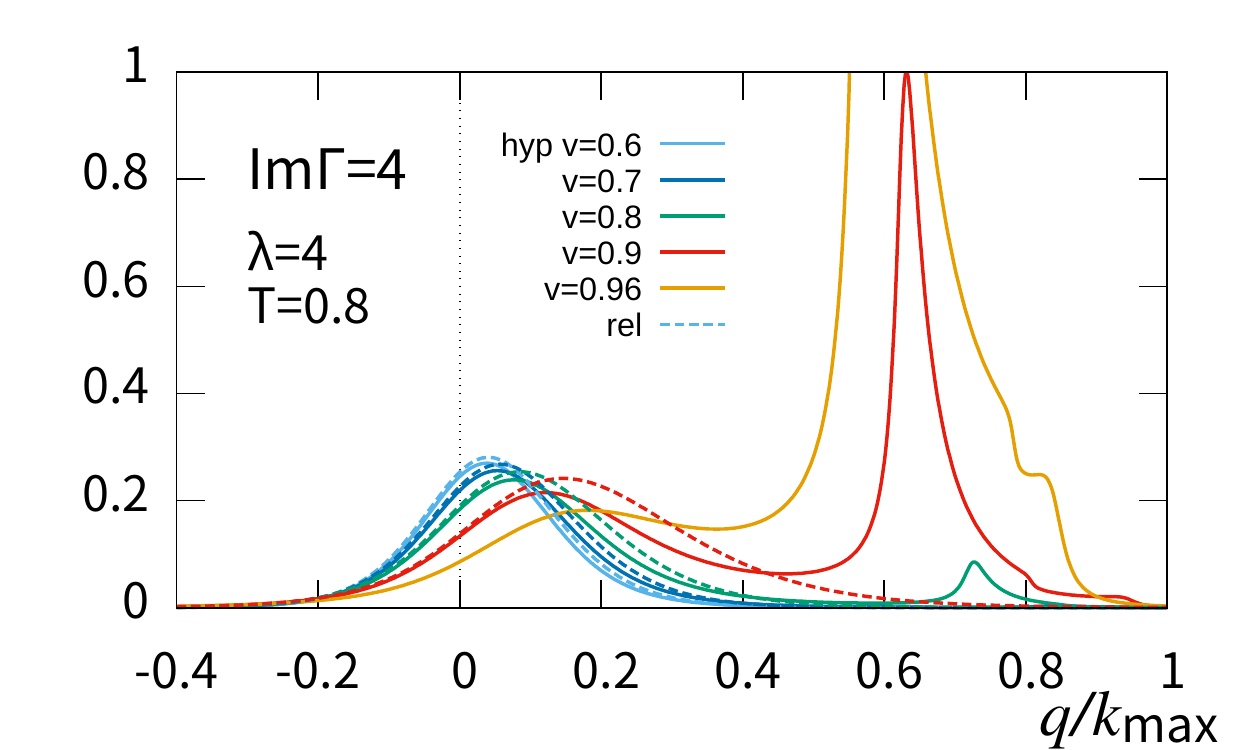}
 \includegraphics[width=0.49\hsize]{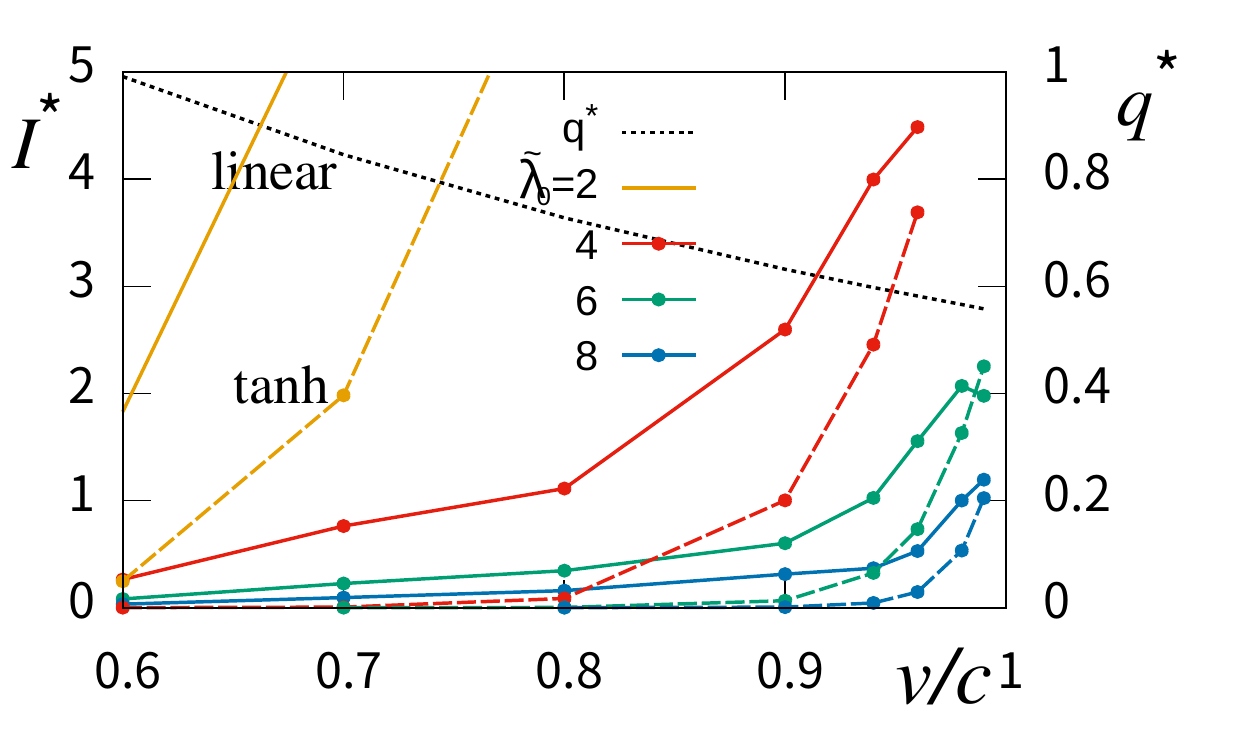}
 \caption{
 (a) $\Im\Gamma_q$ for relativistic (dotted line) and hyperbolic (solid line) dispersion with $\mu=5$ for $\lambda_0/a=4$.
 (b) The peak amplitude $I^*$ and position $q^{*}$ of $\Im\Gamma_q$ for different $\lambda_0/a$. Wall profiles are linear (solid line) and tanh (dashed line).  
 \label{FIGvqI}}
\end{figure}

As Fig. \ref{FIGenergyconserv}(b) suggests, the anomalous emission is enhanced for a band with smaller average slope keeping the maximum slope (the maximum group velocity) as $c$.
We take here as an example a hyperbolic form of 
\begin{align}
{\omega}^{\rm (h)}_k = {\Delta}+\frac{2ck_{\rm max}}{\mu}\lt(1-\frac{1}{\cosh \mu {k}/k_{\rm max}}\rt),  
\end{align}
where $k_{\rm max}=\pi/a$ and $\mu$ is a parameter defining the average slope. \cite{AFSW20dispersion} 
The dispersion does not bring qualitative change in the normal response function $\Pi_q$ (See Ref. \cite{AFSWSM20}), while the imaginary part of the anomalous response $\Im[\Gamma_q]$ is significantly altered (Fig. \ref{FIGvqI}(a)); 
A sharp peak appears for velocity $\Vw/c\gtrsim0.7$ at $q=q^*$ in high $q$-regime ($ 0.5\lesssim q^*/k_{\rm max}\lesssim 1$), indicating strong forward emission of two magnons. 
The minimum velocity necessary is determined by the dispersion; It is obviously larger than 
$\omega_{k_{\rm max}}/k_{\rm max}$ for a monotonically increasing dispersion, which is $\sim \frac{2}{\mu}c$ for hyperbolic dispersion with a small gap.
The peak position $q^*$ is independent on $\lambda_0$. 
The intensity $I^*$ of the peak and $q^*$ are plotted as function of velocity in Fig. \ref{FIGvqI}(b).

The anomalous emission, determined by large $q$ behavior of the response function, is sensitive to the wall profile.  
In the case of very thin wall, linear profile of $n_x$ (or $n_y$) inside the wall may appear instead of the ideal tanh wall, as argued in nanocontacts \cite{TZMG99}. 
For the linear profile, 
$n_x=(1-|x|/\lambda^{\rm (l)})\theta(\lambda^{\rm (l)}-|x|)$, where $\lambda^{\rm (l)}=\lambda \pi/2$, the form factor is
$W_q=\frac{2\pi}{(q\lambda^{\rm (l)})^2}\lt(1-\frac{\sin q\lambda^{\rm (l)}}{q\lambda^{\rm (l)}}\rt)$ (See Ref. \cite{AFSWSM20}).
The anomalous response amplitude $I^*$ is significantly enhanced due to a slower decay at large $q$ (Fig. \ref{FIGvqI}(b)).

The emitted current amplitude is estimated by  
$j\equiv \sum_k q\average{a^\dagger_{-k+q}a^\dagger_{k}}
\sim q^* I^*$.
For $\lambda_0/a=4$, $j\gtrsim 0.8$ for $\Vw/c>0.8$ and $j=0.2$ for $\lambda_0/a=8$ at $\Vw/c=0.9$  at $T=0.8$ and for linear wall profile. 
Let us compare the emitted spin wave current with the current due to the wall motion.
The spin wave current is defined as 
$j=-\frac{i}{2}a^\dagger \stackrel{\leftrightarrow}{\nabla} a
= -\frac{1}{4g}(\dot{\varphi} \stackrel{\leftrightarrow}{\nabla} \varphi) $ 
in terms of real spin wave field $\varphi$.
For a domain wall, 
$ \varphi^{\rm w}=[\cosh \frac{x-X(t)}{\lambda}]^{-1}$.
The current at the wall center is thus 
${ j^{\rm w}} \equiv \frac{V_{\rm w}}{4g\lambda^2} $.
Using $J/a^2\simeq c/a$, we have 
\begin{align}
 { j^{\rm w}} \simeq \frac{k_{\rm max}}{4\pi\tilde{\lambda}_0^2} \frac{\tilde{v}}{1-\tilde{v}^2}
\end{align}
where $\tilde{v}=\Vw/c$.
For $\tilde{\lambda}=4$, 
${ j^{\rm w}} /k_{\rm max}\simeq 0.01(0.02)$ at $\tilde{v}=0.8(0.9)$. 
The current due to the emission is thus by 1-2 orders of magnitude larger than the current of the wall itself in the relativistic regime. 
A thin and relativistic wall is therefore an extremely efficient magnon emitter.

\begin{figure}
 \includegraphics[width=0.49\hsize]{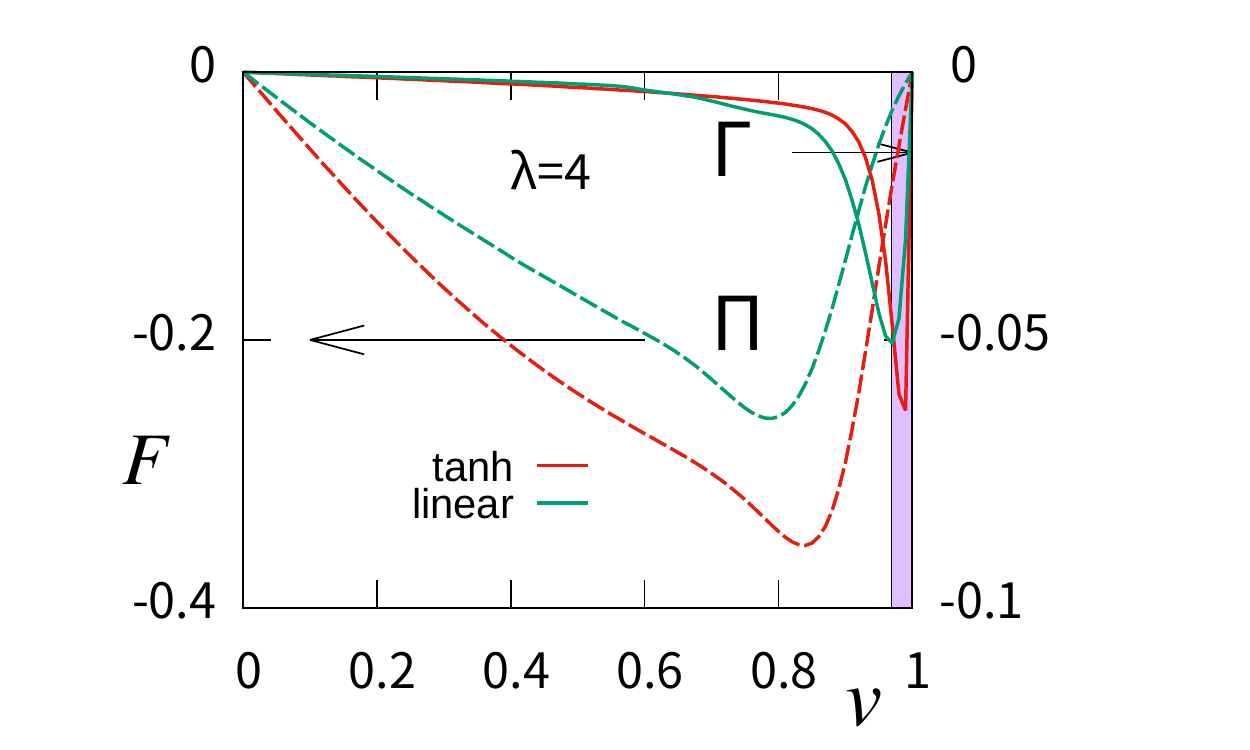}
 \includegraphics[width=0.49\hsize]{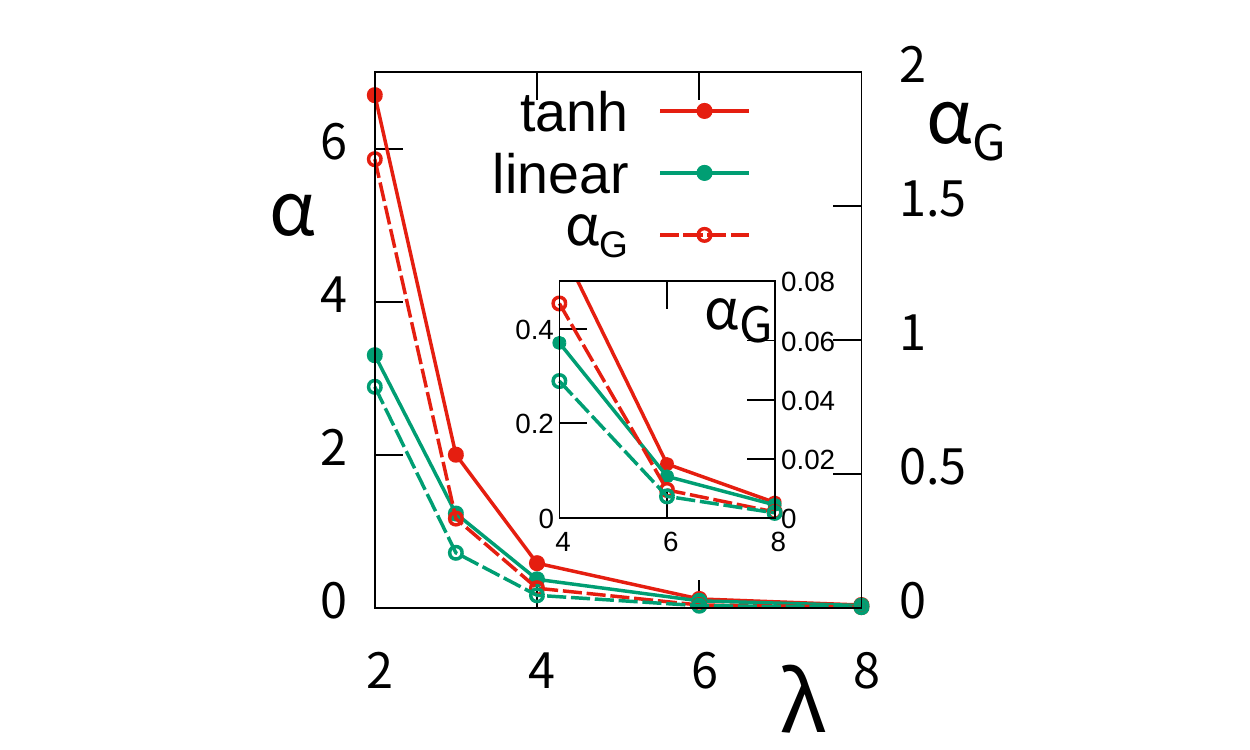}
 \caption{(a) Plot of the force $F$ for tanh and linear wall with $\lambda_0/a=4$. The normal ($\Pi$) and anomalous ($\Gamma$) contributions are shown by dashed and solid lines with axis at the left and right, respectively. Shaded region ($\tilde{v}>0.97$) shows a breakdown of continuum description for $\lambda_0/a=4$. 
 (b) Friction constant $\alpha$ (left axis) and contribution to the Gilbert damping constant (right axis).
\label{FIGF}}
\end{figure}

As reaction to the scattering and emission/absorption, a frictional force, 
\begin{align}
F &=  2\frac{Kg}{a} \Im \sum_{k,q} \frac{W_q e^{-iqX(t)}}{\sqrt{\omega_k\omega_{k+q}}} 
  q\lambda \average{ a_{k+q}^{\dagger} a_{k} +a_{-k}^{\dagger} a_{k+q}^{\dagger} } 
 \label{F}
\end{align}
arises.
As seen in the plot of Fig. \ref{FIGF}, the emission contribution has a narrow peak at high velocity close to 
$\Vw/c=1$, while the normal response ($\Pi$) contribution shows a broad peak starting from low velocity regime.
The normal contribution is larger than the emission contribution as the excitated magnon profile is mostly localized near the wall, resuting in a large overlap. 
The force at small velocity, dominated by the normal response, is an Ohmic friction, 
$F= - \alpha \Vw/\lambda^2$, whose dimensionless coefficient $\alpha$ is plotted in Fig. \ref{FIGF}. 
As the force arises from transfer of finite $q$, the friction constant $\alpha$ depends strongly on the wall thickness. 
The friction coefficient $\alpha$ corresponds to a contribution to the Gilbert damping constant of 
$\alpha_G=\frac{a}{2\lambda}\alpha$, which is plotted by dashed lines.
For linear wall profile, the contribution $\alpha_{\rm G}$ is 0.007 (0.002) for $\lambda/a=6$ (8) at $T=0.8$, which is significantly large compared to the intrinsic Gilbert damping constant of most antiferromagnets. 
The damping due to magnon excitation has clear temperature dependence, exponentially suppressed for $T\lesssim \Delta$ and increases linearly at high temperature below the N\'eel transition temperature \cite{AFSWSM20}. 
For quantitative study, the temperature-dependence of $\eta$ and the fluctuation near the N\'eel temperature need to be taken into account \cite{TataraAF19}.

The direction of the emitted magnons are determined by the sign of the wave vector $k$, while whether it is forward or behind the wall is determined by the group velocity relative to the wall velocity. 
In the case of relativistic dispersion with a gap of $\tilde{\Delta}=0.1$, 
most part of the normal response function at $\Vw=0.8$ turns out to be the magnon excitation behind the wall \cite{AFSWSM20}. 
This is consistent with the observation based on the Landau-Lifshitz-Gilbert (LLG) equation analysis in Ref. \cite{Shiino16} that the moving wall emits magnons mostly backward. 
The LLG study fixes the magnon dispersion to be relativistic, and thus their results are due to the normal response function of the present analysis.

As the amplitude $\average{ a^{\dagger}_{-k+q} a^{\dagger}_{k} }$ indicates, 
the two magnons pair created by the mechanism proposed here are entangled quantum mechanically like in the case of electromagnetism \cite{Ebadi14}, suggesting interesting possibilities for quantum magnonics.

\acknowledgements
This work was supported by 
a Grant-in-Aid for Scientific Research (B) (No. 17H02929) from the Japan Society 
for the Promotion of Science.



\begin{thebibliography}{23}
\bibitem [{AFS({\natexlab{a}})}]{AFSW20Yang}%
  \BibitemOpen
  \href@noop {} {\bibfield  {journal} {\bibinfo  {journal} {Effect of lattice
  beaking the relativistic nature of the wall was discussed in Ref.
  \cite{Yang19}}\ } }\BibitemShut {NoStop}%
\bibitem [{AFS({\natexlab{b}})}]{AFSW20othercollective}%
  \BibitemOpen
  \href@noop {} {\bibfield  {journal} {\bibinfo  {journal} {One could introduce
  other coordinates such as the angle of the wall plane and thickness, as in
  the case of a ferromagnet \cite{TataraSW20}, but they are essentially
  decoupled from the $X$ dynamics, except for the case where the driving field
  mixes them as discussed in Ref. \cite{Shiino16}}\ }
  }\BibitemShut {NoStop}%
\bibitem [{AFS(2020{\natexlab{a}})}]{AFSW20Kim}%
  \BibitemOpen
  \href@noop {} {\bibfield  {journal} {\bibinfo  {journal} {A second-order
  coupling of magnons to a wall velocity was studied in Ref. \cite{KimSW18},
  and it was shown to give rise to a super Ohmic dissipation. Magnon-driven
  domain wall motion was theoretically studied in Ref. \cite{Tveten14}}}
  }\BibitemShut {NoStop}%
\bibitem [{AFS(2020{\natexlab{b}})}]{AFSW20Tveten}%
  \BibitemOpen
  \href@noop {} {\bibfield  {journal} {\bibinfo  {journal} {The potential was
  considered in Ref. \cite{Tveten14} to study magnon-driven motion of
  antiferroamgnetic domain wall}} }\BibitemShut {NoStop}%
\bibitem [{AFS({\natexlab{c}})}]{AFSW20dispersion}%
  \BibitemOpen
  \href@noop {} {\bibfield  {journal} {\bibinfo  {journal} {To be
  mathematically consistent, magnon dispersion is determined by the Lagrangian
  (1) and is correlated with the equation of motion of domain wall. In reality,
  on the other hand, the dispersion is sensitive to details of lattice
  structure, while a collective behavior of a macroscopic wall is not. We thus
  argue the case of a dispersion away from the relativistic limit}}
  }\BibitemShut {NoStop}%
\end{thebibliography}
%

\end{document}